\documentclass[twocolumn,showpacs,preprintnumbers,amsmath,amssymb]{revtex4}
\usepackage{graphicx}% Include figure files
\usepackage{dcolumn}% Align table columns on decimal point
\usepackage{bm}% bold math
\begin{document}
\title{Physics0.01: Object-Oriented Programming for Exact Diagonalization}
\author{MYUNG-HOON CHUNG}

\affiliation{College of Science and Technology, Hong-Ik
University, Chochiwon, Choongnam 339-800, Korea}
\email{mhchung@wow.hongik.ac.kr}

\date{\today}

\begin{abstract}
A new system of library code is proposed and initiated. It is
emphasized that the same terminologies as we find in our textbooks
should be used for class names in the library code. The language
C\# invented by Microsoft is adopted in this project. Several
rules of thumb are suggested in order to obtain easy-readable
coherent codes. As a first step, we present the library code for
exact diagonalization in physics. When we build codes, we clearly
distinguish between model independent and dependent parts, and we
use familiar terminologies like {\sf Hamiltonian}, {\sf
HilbertSpace}, {\sf GroundState}, {\it etc} as class names. As an
explicit example, we calculate ground state energy of a quantum
dot, showing the triplet-singlet transition.
\end{abstract}


\maketitle

\section{Introduction}

In Feb. 2002, Microsoft launched a new program language
named C\#, which is strongly object-oriented\cite{1}. It is argued
that C\# language has three characteristic features, 1. C\# is as
elegance as Java, 2. C\# is as powerful as C++, 3. C\# is as
productive as Visual Basic. Although it is still questionable
whether or not C\# is useful for scientific computing in academic
area, we here adopt C\# because of its strong character of
object-oriented language.

Reusability is always one of main concerns in the area of
software. Scientific computing also should require
reusability seriously. We notice many efforts to improve reusability. The
web site managed by Troyer\cite{2} is remarkable. The library code
for density matrix renormalization group method invented by
White\cite{3} is also well known. These library codes are written
in C++. There are many fortran codes in netlib\cite{4}. It is
clear that there is a general trend of transition from non
object-oriented to object-oriented language.

The main advantage of C\# over C++ is efficiency in making Windows
applications. Since computer users are addicted by Windows
operating system, user friendly programs would be Windows
applications. As far as end users need more user friendly
programs, C\# can be useful. One good point is that so called Mono
project\cite{5} is carried on to develop the C\# compiler for
Linux. After successful finish of Mono project, we guess that C\#
will become more popular.

In this article, we propose a new way of making library codes with
C\# as a preparation for the battle between C\# and C++, which
will take place soon. At present, C++ is the most powerful.
However, after someone makes ease access to parallel computing in
loosely connected Windows operating systems using XML web
services, we expect that C\# can compete with C++ even in
scientific computing.

\section{Supercode}

In order to improve the power of scientific computing, we have to
consider the three factors: CPU time, REM memory, and Coding time.
For some complicated problems, the main factor is nothing but
coding time. In this situation, someone want to use Standardized
Reusable Components (SRC). When we say {\sf Hamiltonian}, everyone
in physics community can understand the meaning of it. Hence,
there must be a class called {\sf Hamiltonian}. At this moment, we should
notice the usefulness, when we use the same terminology as we find
in our textbooks. All library codes must be organized like our
textbooks and our library. The system of library codes should not
be in simple alphabet order, but would be in subject-based order.
This subject-based integration of library codes will result in high
correlations between codes, in other words, it will need more
coherence between codes. It is not an easy task to make
classification in each category of our knowledge. We may need a
big discussion on this matter. At any rate, this library codes can
be called {\bf Supercode}. As we see in
usual real library, we can divide {\bf Supercode} as {\bf Biology},
{\bf Chemistry}, {\bf ElectricalEngineering}, {\bf Mathematics},
{\bf MechanicalEngineering}, {\bf Physics}, {\it etc}.

In scientific computing, programmers always want to know results
as quickly as possible. Because of this hurry habit, program codes
become less structural, and reusability is lost. In order to
enhance reusability, it is strongly proposed that scientists
should make program codes in distinguishing model-independent
and model-dependent parts. The model-independent parts are
properly organized and will be upgraded in the future. Numeric
workers should be familiar with the whole structure of
model-independent parts, and should handle model-independent and
model-dependent parts simultaneously. In order to emphasize the
way of coding, we denote HANA, which is abbreviation for ``Handle
All aNd All", where the first ``All" means model-independent parts
and the second ``All" means model-dependent parts. We define
HANA project as efforts to make {\bf Supercode}. We are
summing up some rules of thumb for {\bf Supercode}:

1. {\em Divide and Conquer}. Do not exceed the total number of
code over 100 lines for each class.

2. {\em Handle all and all}. Consider always a general code for a
general case as much as possible. Collect the codes of
model-independent parts for {\bf Supercode}.

3. {\em Give proper full names}. Be careful in choosing namespace
names and class names. For instance, all codes in Numerical
Recipes should become {\tt private} methods in {\bf Supercode}. An
easy way to choose names is to use an index in the end pages of
textbooks. Allow no space in a full name, instead use upper case
characters, for instance, {\bf ProteinFolding}, {\bf
SpecialFunction}, {\bf QuantumMechanics}, {\it etc}. This rule was
adopted in the textbook written by Deitel, ``C\# How to program".

4. {\em Use SRC}. Use confirmed codes as much as possible. In
order to avoid any mistakes, do not use special or home-made
codes.

5. {\em Reduce number of arguments in constructors and {\tt
public} methods}. Future library users will definitely want a
simple structure of arguments.

6. {\em Consider polymorphism in the pattern of ``word after the
same word"}. Make an abstract class for the common name. For
instance, when we note Translation Symmetry, Rotation Symmetry,
Reflection Symmetry, Spin Inversion Symmetry, and so on, we make
abstract class {\sf Symmetry} and subdirectory {\bf Symmetry}, and
then make class {\sf Translation}, {\sf Rotation}, {\sf
Reflection}, {\sf SpinInversion} in {\bf Symmetry} directory,
using the key word {\tt override}.

Because of the author's speciality, only a small part of {\bf
Supercode} is discussed in the following. The part is named as
{\bf Physics0.01}. The version number 0.01 represents the first
step toward the final goal of {\bf Supercode}. We hope that
upgrade version of {\bf Physics} will be used for all computing
processes in physics.

\section{Physics0.01}

Following the above rules, we try to construct {\bf Physics0.01}.
All files related to physics exist in a single directory named
{\bf Physics0.01}, which contains many subdirectories and files.
Here the subdirectories are names of namespaces, and files are
names of classes. The subdirectories, which can be called volumes,
are classified as usual as textbooks for physics. We make the
subdirectories: {\bf ClassicalMechanics}, {\bf QuantumMechanics},
{\bf Electromagnetism}, {\bf StatisticalMechanics}, {\bf
GeneralPhysics}, {\it etc}. One of these volumes, {\bf
QuantumMechanics}, has many class files: {\sf
ConservedQuantity.cs}, {\sf GroundState.cs}, {\it etc}, and
subdirectories: {\bf Hamiltonian}, {\bf HilbertSpace}, {\bf
Symmetry}, {\bf Operator}. The directory {\bf Hamiltonian}
contains many classes. Following the sixth rule of polymorphism,
the class files in {\bf Hamiltonian} directory are given by
{\sf FermionOneBody.cs}, {\sf FermionTwoBody.cs}, {\sf
Heisenberg.cs}, {\sf Kondo.cs}, and abstract class {\sf
Hamiltonian.cs}, since we note Fermion One Body Hamiltonian,
Fermion Two Body Hamiltonian, Heisenberg Hamiltonian, and Kondo
Hamiltonian. Still many other class codes should be made,
accumulated and organized in {\bf QuantumMechanics} of {\bf
Physics0.01}.

We here omit to show the present codes of {\bf Physics0.01}. All
detailed codes are freely open in the author's home page\cite{6}.
This prototype object-oriented programming for physics is invented
mainly only for exact diagonalization. We expect that Monte Carlo
and Maxwell equation will be included. Hence upgrade {\bf
Physics0.02} will appear soon.

\section{Example: Ground State Energy in 2D Parabolic Quantum Dot}

In order to present the general procedure of HANA project, we
consider the problem of evaluating ground state energy in a
two-dimensional parabolic quantum dot. In this problem, the second
quantized Hamiltonian\cite{7} is written as
\begin{eqnarray}
H &=&\sum_{n,l,\sigma} (\hbar\omega(2n+|l|+1)
-\frac{1}{2}\hbar\omega_{\mbox{c}}l - g
\frac{1}{2}\hbar\omega_{\mbox{c}}\sigma)
c_{nl\sigma}^{\dagger}c_{nl\sigma}  \nonumber \\
&+&\frac{1}{2}\sum_{n_{1},n_{2},n_{3},n_{4},k,l,m}
V_{n_{1}n_{2}n_{3}n_{4}}(k,l;m)  \nonumber \\
&\times&\sum_{\sigma,\sigma^{\prime}}
c_{n_{1}k\sigma}^{\dagger}c_{n_{2}l+m\sigma^{\prime}}^{\dagger}
c_{n_{3}l\sigma^{\prime}}c_{n_{4}k+m\sigma},
\end{eqnarray}
where one-particle creation operators $c_{nl\sigma}^{\dagger}$
have three quantum numbers\cite{8}. The principal quantum number
$n$ runs as $0$, $1$, $2$, $3$, $\cdots$, while the angular
momentum quantum number $l$ is given by $0,~\pm 1,~\pm 2,~\pm
3,~\cdots$, and the spin index $\sigma=\pm 1$. Note that the
confining potential is related to $\omega_{0}$ in the usual way,
and $\omega
=\sqrt{\omega_{0}^{2}+\frac{1}{4}\omega_{\mbox{c}}^{2}}$, where
the so called cyclotron frequency is proportional to the strength
of magnetic field $B$ as $\omega_{\mbox{c}}=eB/m^{\ast}c$. It is
straightforward to calculate the Coulomb matrix elements:
\begin{eqnarray}
\frac{1}{2}V_{n_{1}n_{2}n_{3}n_{4}}(k,l;m)
=\frac{1}{2}\frac{e^{2}}{\kappa}\sqrt{\frac{m^{\ast}\omega}{2\hbar}}
C_{n_{1}n_{2}n_{3}n_{4}}(k,l;m) \nonumber \\
~~=\hbar \omega
\frac{58.47}{\kappa}\sqrt{\frac{m^{\ast}}{m}}\sqrt{\frac{1\mbox{meV}}{\hbar
\omega }}C_{n_{1}n_{2}n_{3}n_{4}}(k,l;m),
\end{eqnarray}
where $\kappa$ is a dielectric constant, and
$C_{n_{1}n_{2}n_{3}n_{4}}(k,l;m)$ are dimensionless numbers.

With the usual bulk values, $m^{\ast}=0.07 m$ and $\kappa =13$,
the overall factor of the Coulomb interaction becomes
$58.47/13*\sqrt{0.07}=1.19$. However, since the used quantum dot
looks two-dimensional, it seems that the values of dielectric
constant and effective mass are not known. Thus, it is needed to take
the overall factor as an input parameter. The value of
$\omega_{0}$ is also an input parameter. In fact,
the side gate is used in order to control the confining potential\cite{9}. In
consequence, this calculation has four input parameters: number of
electrons, the overall factor, $\omega_{0}$, and
$\omega_{\mbox{c}}$. We take 20 lowest energy states as $2n+|l|+1 \le
4$ in this exact diagonalization. Only the main parts of the code
are presented as follows.
\begin{widetext}

$\cdots\cdots$
\begin{verbatim}
 using QM=Physics.QuantumMechanics;
 using H=Physics.QuantumMechanics.Hamiltonian;
 using HS=Physics.QuantumMechanics.HilbertSpace;
 using OP=Physics.QuantumMechanics.Operator;
\end{verbatim}

$\cdots\cdots$
\begin{verbatim}
 int Ne = Convert.ToInt32(textBox2.Text); \\ input number of electrons
 double factor = Convert.ToDouble(textBox3.Text); \\ input the overall factor
 double omega0 = Convert.ToDouble(textBox4.Text); \\ input omega0
 double omegac = Convert.ToDouble(textBox5.Text); \\ input omegac

 int half = (numberOfElectrons + (numberOfElectrons % 2))/2; \\ At half, S_{z}=0

 QM.Space i1 = new QM.Space(7,2); \\ 14 sites
 QM.Space i2 = new QM.Space(3,2); \\ 6 sites
 QM.Space ii = new QM.Space(i1,i2); \\add Space i1 and i2, so 20 states

 QM.QuantumNumber qn = new QuantumNumberForModel(ii); \\assign n, l and spin
 QM.Coupling u = new CouplingForModel0(qn,omega0,omegac); \\one body spectrum
 QM.Coupling w = new CouplingForModel1(qn,factor,omega0,omegac); \\Coulomb matrix

 OP.FermionCreation cdagger = new OP.FermionCreation(ii);
 OP.FermionAnnihilation c = new OP.FermionAnnihilation(ii);

 H.Hamiltonian h0 = new H.FermionOneBody(u, cdagger, c);
 H.Hamiltonian h1 = new H.FermionTwoBody(w, cdagger, cdagger, c, c);

 H.TotalHamiltonian h = new H.TotalHamiltonian(h0,h1);
 h.ZeroPointEnergy = 0.0;

 OP.Number nop = new OP.Number(numberOfElectrons);
 for(int ta = 0; ta < 3; ta++){
 for(int ts = half; ts < half+2; ts++){
 QM.ConservedQuantity totalAngularMomentum = new QM.ConservedQuantity(qn,1,ta);
 QM.ConservedQuantity totalSpin = new QM.ConservedQuantity(qn,2,ts);

 HS.HilbertSpace hil1 = new HS.Fermion(cdagger, nop);
 HS.HilbertSpace hil2 = new HS.Reduction(hil1,totalAngularMomentum);
 HS.HilbertSpace hil = new HS.Reduction(hil2,totalSpin);

 QM.GroundState psi = new QM.GroundState(h,hil);
 textBox1.Text = Convert.ToString(psi.Energy);}}
\end{verbatim}

$\cdots\cdots$

\end{widetext}

Note that the classes of model-independent parts are in specific
directories. The model-dependent parts consist of {\sf
QuantumDot.cs}, {\sf QuantumNumberForModel.cs}, {\sf
CouplingForModel0.cs} and {\sf CouplingForModel1.cs}. The user
friendly simulation program runs in Windows operating system.
Giving the four input parameters, we simply click a button to find
the ground state energy. The final result of ground state energy
is printed on the screen.

\begin{table}
\caption{\label{tab:table1}The ground state energy in each sector.
We notice that there is the triplet-singlet transition between 1.1
and 1.2.}
\begin{ruledtabular}
\begin{tabular}{ccccc}
 \multicolumn{2}{c}{Sector} &  \multicolumn{3}{c}{Ground State
 Energy}\\
$S_{z}$&$L_{z}$&$\hbar \omega_{0}=1.1$&$\hbar \omega_{0}=1.15$&$\hbar \omega_{0}=1.2$\\
\hline
 0    &    0    &     59.7012     & 61.4422     & 63.1693\\
 0    &    1    &     59.7247     & 61.4975     & 63.2490\\
 1    &    0    &     59.7012     & 61.4759     & 63.2299\\
 1    &    1    &     59.7247     & 61.4975     & 63.2490\\
\end{tabular}
\end{ruledtabular}
\end{table}

We present the result of calculation related to the
triplet-singlet transition\cite{10}, when the strength of
confining potential is increased in the dot containing six
electrons. We let the overall dimensionless factor be 2, and
$\omega_{\mbox{c}}=0$. We find the transition near $\hbar
\omega_{0}=1.14$ meV as shown in Table~\ref{tab:table1}. We expect
that someone observe this transition using the side gate voltage
in the future.

\section{Conclusion}

We proposed a library code system, emphasizing that class names
should be the same as we find in textbooks. Like an well known
idiom in algorithm, ``Divide and conquer", we introduced ``Handle
all and all" for this library system. The first ``all" means
library codes and the second ``all" means our specific project
codes. Hence, there are only two folders: one is library and the
other is our specific model dependent project. As the library grows,
we expect that each class becomes strongly correlated with others.
To change codes is not an easy task because of coherence between
codes. However, experts can freely change the codes, and thus
invent the upgrade version with fixing the first number such as
{\bf Physics1.03} from the mother version {\bf Physics1.00}. For
instance, if someone feel that the class {\sf Symmetry} can be
improved, then change the code. If this change is worthwhile for
everyone, then it should be reflected in the next version of {\b Physics} with
clear notification of contributors. Along this line, the prototype
library code, {\bf Physics0.01}, will be upgraded soon.

We know that nowadays many theoretical physicists spend more time
in complicated computing, which is however conceptually trivial.
We agree that physicists need to do philosophical and conceptual
thinking more. The author likes to imagine that a higher version
of {\bf Physics} will reduce the time of computing for all
physicists.

\section*{Acknowledgment}

The author learned C\# language when he stayed at Yale University.
He gratefully thanks to Professor Paul Hudak, who did an
excellent job in his C\# class.

\end{document}